# Positron Annihilation Study of Zn O Nanoparticles Grown under Folic Acid Template

**Sreetama Dutta, Sourav Sarkar[§] and Bichitra Nandi Ganguly[1]**
Saha Institute of Nuclear Physics, 1/AF Bidhannagar, KOLKATA, INDIA.,
[§] IISER,KOLKATA, INDIA

[1]Corresponding author's email: bichitra.ganguly@saha.ac.in

**Abstract**. Positron lifetime spectroscopy (PAL), Doppler broadening (DB) as well as coincidence Doppler broadening (CDB) spectroscopy of a new variety of Folic acid (FA) capped zinc oxide nano-particle samples has been performed at room temperature. The results show interesting patterns of observation, hither-to unobserved in ZnO wurtzite crystalline samples, such as predominance of positronium formation, as reflected in the (PAL) analysis, phase transition in the nano crystalline samples at ~0.8-1.0 % FA concentration as depicted from DB results. Also, the chemical environment of the samples has been analysed from the ratio curves of CDB studies. Beside these, other independent results from X-ray diffraction (XRD) data and Debye- Scherrer method, Transmission electron microscopic (TEM) observations and as well as Fourier transformed infrared spectroscopic (FT-IR) analysis are reported for comparison.

## 1. Introduction

Despite manifold interests on the wide band gap (3.37 eV) semiconductor properties of zinc oxide (ZnO) material and its varied optoelectronic applications[1,2], much remains to be explored when its material perspective is sought in the context of nanotechnology and biomedicine. ZnO is generally known to be nontoxic and biocompatible substance[3]. It is also realized that ZnO nanomaterials with the size dependent surface properties hold considerable promise as anticancer agent; the area is sensitive and important to biomedicine[4]. The field eventually leads to versatile platforms for biomedical applications and therapeutic intervention.

Growing nano sized mono dispersed ZnO has been one of the very challenging tasks, which we have developed in our laboratory using sol-gel technique, under the template of a biologically important molecule such as folic acid under physiological conditions, maintaining the pH of the medium and temperature[5]. The characteristic physical properties of the nano sized material so prepared under the bio-template has been examined initially with the help of various physical techniques like XRD analysis, FT-IR spectroscopy, for its purity and chemical bonding aspects[5].

But for the detail micro structural analysis and any surface vacancy related studies in the wurtzite micro crystalline ZnO nanomaterials, that might reveal the exact mechanism of surface related interaction or any residual charge effects which may vary under the preparation condition with respect to the particle size effects and folic acid concentration, the material has been subjected to positron annihilation spectroscopic studies (PAS), using positron life time (PAL) measurements of the samples and the Doppler broadening (DB) line shape studies with 'S' and 'W' parameter analysis. A coincidence Doppler broadening (CDB) study has been further under taken, with the help of a second (highly efficient NaI (Tl) scintillator) detector in coincidence to HPGe solid state detector to minimize the back ground effects, in order to perceive the differences with highly pure sol-gel grown ZnO oxide nanoparticle (wurtzite structure) sample and the bio-templated sol-gel grown sample, with respect to momentum distribution analysis (at the higher momentum side, 511keV+$\Delta$E), considering the ratio curves of the CDB, which significantly attributes to the ensconced chemical environment[6] of the nano-particles. The results reflect interesting transition in molecular properties from the DB line shape analysis and an intriguing features in the structural properties *vis-a vis* the analysis from PAL and CDB due to capping of the nano particles within folic acid environment. The effects are in corroboration with other complementary techniques like XRD and Debye-Scherrer method, FT-IR analysis and TEM observation which affirms the changes in structural morphology as well. Although ZnO nano particles have been investigated by many investigators earlier[7], but study of bio-functionalized nano particles, by PAS is sparse. The use of PAS as a micro-structural probe, for elucidation of modifications in the structural properties of ZnO-folic acid conjugate opens a new avenue as has been delineated in this article.

## 2. Experiment

### *2.1. Sample preparation:*

ZnO pure nano-particles were prepared by the sol-gel technique[5] from zinc acetate (A.R. grade), in alkaline solution phase using 1:1 ammonia solution (Merck, India), at pH=7.5. The Folic acid templated sample has been obtained using Folic acid (M.F: $C_{19}H_{19}N_7O_6$ procured from Sigma Life Science) in the same alkaline solution at room temperature. In all cases of solution preparation, triple distilled water has been used.

### 2.2. *X-ray diffraction study:*

The phase structures of the samples were identified by X-ray diffraction technique using Rigaku TTRAX3 diffractometer with CuKα radiation (wavelength of the radiation, k = 1.54 Å). The data have been collected in the range (2θ) $20^o$ –$80^o$ with a step size of $0.02^o$. Si has been used as external standard to deconvolute the contribution of instrumental broadening. The grain sizes of the synthesized samples have been calculated using Scherrer formula [8]: $D_{hkl} = K\lambda/\beta_P \cos\theta$, where, $D_{hkl}$ is the average grain size, K the shape factor (taken as 0.9), λ is the X-ray wavelength, $\beta_P$ is the full width at half maximum (FWHM) intensity and θ is the Bragg angle. The results are shown in Fig 1. and Table1.

### 2.2. *Transmission Electron Microscopic (TEM) study:*

The morphology of the synthesized product were characterized by transmission electron microscopy, TEM (Tecnai S-twin, FEI) using an accelerating voltage of 200 kV, having a resolution of ~ 1 Å. The nano crystalline grains were dispersed in isopropyl alcohol and sonicated to remove any agglomerated particles and dried under high vacuum prior to examination under TEM. The results are shown in Fig. 2. and schematically explained in Fig 3.

**2.3. *Positron Annihilation spectroscopy*:**

**a) Lifetime measurements**: A fast–fast gamma-gamma coincidence circuit was used for the purpose of measuring positron annihilation lifetime spectrum using spectroscopic quality $BaF_2$ as the scintillators (cone-shaped) coupled to XP2020Q photomultiplier tubes as the detectors, with the positron source($Na^{22}$ in the form of NaCl, ~5 μ Ci point source, on an ultra thin Ni foil~3μm and folded) and sandwiched between the compact samples (in pallet form). A total of $10^6$ counts were recorded under the spectrum with a peak to back ground ratio 5000:1. The resolution of the spectrometer was ~260 ps, as measured by $Co^{60}$ source at the positron window settings. The spectrum was analyzed by latest version of PATFIT package program[9]namely PALSfit, with a necessary source correction which showed three life time components , the results are shown in Fig. 4 and Fig. 5.

**b) Doppler broadening (line shape) spectroscopy:** The detector used in DB spectrometer is a close ended co-axial high purity Ge (HPGe) detector(p-type) from ORTEC. Standard sources namely $Co^{60}$ with γ photopeak at 1.33 MeV, the energy resolution (FWHM ~1.75 keV), and with $Bi^{207}$ for the γ photopeak at 579 keV with the energy resolution (FWHM~ 1.1keV) were used for calibration. The experimental spectrum was recorded around 511 keV annihilation line in order to determine the 'S' parameter, this being evaluated as the ratio of counts in the central region (511±1.7 keV) of the spectrum to the total counts recorded under the spectrum. The results are represented in Fig. 6. Also 'W' parameter was measured at the wings (2keV< |E-511| <7keV) for the different samples on either side of the Doppler spectrum to the total fixed counts under the spectrum. The ratio of the S-W plot is presented in Fig. 7.

**c) Coincidence Doppler broadening spectroscopy:** This technique essentially uses another detector kept at 180º with the direction of the gamma rays by the original HPGe detector, so that the other gamma ray is detected in the second detector (in this case 3"x3"NaI(Tl) scintillator is used for its good efficiency) in coincidence and is used to gate a gated biased amplifier through which the amplifier output signals are fed to the multi channel analyser(see Fig. 8). This arrangement essentially reduces the background at the higher energy side and thus improves the peak to background ratio (5000:1), also reduces the energy resolution by a factor ~√2 such that at the wings especially at the right hand side of the spectrum (511keV+ΔE). The higher momentum distribution is perceived very clearly[10] after back ground subtraction and peak normalization with the reference standard sample taken as pure ZnO . The ratio curves of ZnO-FA /pure ZnO were generated, as shown in Fig. 9.

**2.5. *FT-IR Spectroscopic measurements :***

Fourier transmission infrared (FT-IR) spectra of the powders (as pellets in KBr, without moisture) were recorded using a Fourier transform infrared spectrometer (Perkin Elmer FTIR system; Spectrum GX) in the range of 400-6000 $cm^{-1}$ with a resolution of 0.2 $cm^{-1}$. The results are shown in Fig. 10.

## 3. Result and Discussion

***Size and morphology*** : Sol-gel grown ZnO oxide nano particles offer high grade purity of the samples as well as nearly monodispersed particles (controlled size distribution) with only proper treatment with the capping agent, the excess of the same could be washed –off at the end. The result of such a synthesized sample is presented through Fig.1 and Table 1. and through the TEM micrographs in Fig 2. Although the physical characteristics and the wurtzite structure of the pure ZnO nano crystalline samples and the FA capped ZnO nano particles have been documented[5], however the size of these nano particles vary according to the concentration of the FA used in the medium, the time taken for growing and isolation of the particles while under preparation. A partial crystallinity[5]was observed with increase in concentration of FA to 4.8%[5]. The purpose of addition of FA in extremely small amounts was to control the size of the nano particles and to achieve monodispersity of the synthesized particles. It has been found that freshly prepared solution and immediate isolation of the as grown particles offer better results. The sonication of the particles and drying immediately on the microscopic carbon coated grids under high vacuum showed the particles to be less than 10 nm. In the electron micrograph, in average particle size of $\leq 10$ nm can be observed for pure ZnO. But the FA ensconced ZnO nano particles show although a swollen structure ~11nm, and the particles are properly isolated. An illustration of the bare nano particles and FA templated nano particles can be found in Fig. 3. The excess addition of FA show an encapsulated spindle shaped structure which are much greater in size. There is however a variation in size, as shown in Table 1 as compared to Fig.2. This could be due to sonication and the drying processes involved and the inherent variations of the detection methods.

***Micro structural analysis*** **:** In order to investigate the detail micro structural analysis and any surface vacancy related studies in the wurtzite nano-crystalline ZnO grains, that might reveal the exact mechanism of surface interaction or any residual charge effects which may vary under the preparation condition, with respect to the particle size effects and folic acid concentration, the material has been subjected to positron annihilation studies using PAS, DB as well as CDB method. The qualitative variation in the positron life time parameters in the samples denotes the variation of defect related properties or any structural variation in the material as positron parameters are sensitive to any kind of electronic microenvironment. In the case of FA templated ZnO nano particles, results of the positron life time analysis could be fitted to three components, along with their respective intensities from the three different electronic states, including the case of pure ZnO sample (standard), which is has been shown in Fig 4.and Fig.5.

The shortest life time is represented by $\tau_1$. The bulk life time[11] of positrons, $\boldsymbol{\tau_b}$ , of the pure ZnO sample has been calculated to be ~ 153 ps and is in accordance with the literature[12] value. However, as the FA percentage increases in the system, the $\tau_1$ shows a change in the trend, since at this stage of preparation, due to capping, the surface state and trapping rates become important. With ~1% FA, a transition has been noticed. Initially since $\tau_1 < \tau_b$ (~0.153 ns), a fraction of positrons could be trapped with negatively charged $Zn^{+2}$ monovacancy defects[13]. The trapped positrons' life time is expected to be larger than $\boldsymbol{\tau_b}$, depending up on the electron density at that defect site. The trapping rate is however found to be very weak, and of the order as: $\boldsymbol{k = 1.8\ (ns)^{-1}}$; with effective annihilation rate as $\lambda_{eff} = \mathbf{8.33\ (ns)^{-1}}$. Assuming a diffusion coefficient ~ $\mathbf{10^{-5}cm^2S^{-1}}$ for positrons[14], there by the diffusion length was found to be **≤ 10 nm.,** which could be just at grain boundary or surface region, or even diffuse out towards low electron density region coronated by FA as shown in Fig 3.when the actual ZnO grain sizes are smaller.

The corresponding $I_1$ percentages (Fig. 5a) from this state have been found to increase, with FA in the system, as more and more of the positrons could be trapped at this site, either for neutralizing the defects or it is possible that interaction of FA as electron donor at the nano grain surface enhances such a possibility, where positrons could be drawn. Thus, effectively we find surface charge effect has been noticeable.

The second life time component $\tau_2$, varies between 370- 470ps (Fig 4b), could be assigned to higher order defects at the surface or interfacial region(Fig 3b), depending up on the structural condition as the morphology varies within the system. In the nano sized particles the surface/volume ratio increases and in general the molecules are loosely bound, therefore annihilation at the surface/interface region could give rise to increase in the $\tau_2$ values. The concentration of FA at the interface at higher concentration (at 4.8% FA) shows a drastic decreasing trend due to close packing, this is also discernible from the micrograph Fig. 2c.

$I_2$ percentage shows more or less uniform pattern within (Fig.5b )the FA conjugated structure, A levelling-off of the same is due to the strongly ensconced nature of the nano particles as the FA concentration increases in the system(Fig 4b). This also corroborates with the morphological changes as observed by TEM Fig 2.

A profound effect is that of positronium (Ps) formation in the system that is reflected by the third long life time component (in ns) Fig 4c and 5c, which is a dominant feature of the system, it is greater than 5% , wherever FA has been used, since free volume structure increases with addition FA in the system. These results also commensurate with the change of the morphology of the system, studied by Fig 2. and the illustration of the same in Fig.3. Intensity shows a decreasing trend at higher concentration of FA. This could be simply due to density effect of FA, i.e. increase in the number of molecules of FA that results in the strongly ensconced structure of the nano particles (spindle shaped structure in Fig 2) as percentage of the same increases at the preparative stages.

The momentum distribution of the annihilation radiation has been shown by the DB results in Fig.6. A structural transition in the nano particle system has been observed at 0.8-1% FA capped system from DB results through S- parameter. As the surface structure gets modified (S-parameter is sensitive to valance electron states), substrate is comparatively less densely packed as FA concentration gradually reaches a critical value, S parameter increases with the open volume defect enriched state (compare with Fig.4., Fig.5. of the positron life time results) but further with increase in concentration of FA the effect levels-off. This observation of abrupt change in line shape parameter denotes a structural transition which is also in accordance with our earlier results [5] as also the same is reflected with the $\tau_1$values in PAL. W–parameter is also measured, the S/W ratio shows a linear fitting curve Fig 7, denoting a uniform trend of defect structure, but as shown by the arrow, the transition point stands alone at the top, denoting a different structural characteristic feature and the encircled points are actually grouped near about the transition point.

At the high momentum wings of the DB line shape analysis, because of the low background effect due to CDB system with an additional detector vide Fig 8, there exists the characteristic signature of the chemical environment due to capping effect of FA. This effect can be best perceived by the ratio curves of CDB, when the back ground effect is minimized, Fig.9. shows the signature of chemical environment due to FA through 2s,2p electrons of carbon and nitrogen in the co-ordinating groups like–COOH and -NH2 (~ I.P. ~9.4 eV[14], binding energies ~ 10 eV that could correspond to a DB energy of 2.5 - 3keV), the signature of which towards the right wing side of the spectrum (511keV+ ΔE) can be significantly discernible from the improved energy resolution of the detector and due to very low back ground )[10]. Really, at the high momentum side the signature of the core momentum, arising in the range of ~35-40 $x10^{-3}$ $m_0C$ (due to 1S electrons of carbon) has been found at the far end, as the signature of core electrons from the ligands of the organic functional (C=O) group ornamenting the ZnO nano particles[14].

Concomitantly, presence of such chemical environment at the surface of the nano particles prepared in the template of FA could be also decided almost independently from the IR spectroscopic study of the material as shown in Fig 10., Table II. This characteristic emblem of FA is very important as the same makes ZnO nano particles bio-chemically important for targeted drug delivery purposes[4]. In this aspect, positron annihilation parameters have been shown to serve as a surface sensitive probe for the assay of the micro structural aspects of nano particles.

## 4. Conclusion.

Study of ZnO nano particles of size less than 10 nm is feasible through positron annihilation spectroscopy, that senses the defects mostly at the surface and brings up the nature of surface involved in the template, along with the elemental features of the coordinating chemical group, clearly discernible from CDB studies. There is an onset of a structural transition in the presence of around 1% FA in the preparation system of the ZnO nano grains as is indicated in DB studies.

**Acknowledgement :** The authors acknowledge the technical help received from **Soma Roy** for all spectroscopic data recording and **Pulak Roy** for TEM pictures.

**6.Tables :**

**Table 1**. The Size variation of pure and FA templated ZnO samples as measured by Debye- Scherrer method from XRD data.

| Samples | Size (nm) |
|---|---|
| Pure ZnO | ≤ 10 |
| 0.5% FA | 18 ±1 |
| 1.3% FA | 18 ±1 |
| 2.0 % FA | 20 ± 2 |
| 4.8 % FA | 20 ± 2 |

**Table II.** The characteristic IR frequencies ($cm^{-1}$) of FA as observed with the ZnO nano particles.

| Wave number ($cm^{-1}$) | Absorption band |
|---|---|
| ~ 3400-3600 | stretching O-H mode, N-H mode |
| ~ 1600 | symmetric C=O stretching mode,<br>stretching $CONH_2$, N-H bending vibration |
| ~ 1380 | asymmetric C=O stretching mode |
| ~ 500 | Zn-O stretching mode |

**7. Captions to the Figures**:

1. Chracteristic X-ray diffraction pattern of the synthesized nano grains of pure-ZnO as a representative example for calculating the grain size[5].

2. TEM picture representing a) Pure ZnO nano grains , b) folic acid ensconced ZnO-FA(2%) showing swollen structure and c)in presence of high concentration of folic acid, ZnO-FA(4.8%) showing spindle shaped structure.

3. Schematic diagram of the a) pure ZnO nano grains b) ZnO nano grains in coronated by the presence of the organic functional groups such as folic acid (FA). c) showing loosely bound structure of the ZnO nano particles and FA with free space.

4. Variation of positron life time results in ZnO nano grains in the presence of the variation of folic acid (FA%) ornamented structure , a. ($\tau_1$), b. ($\tau_2$), c. ($\tau_3$).

5. Variation of relative percentage intensities of different positron life time components : a. $I_{1,}$ b. $I_2$, c. $I_{3.}$ in the presence of variation of FA % in the ZnO nano crystallites preparation stages.

6. Doppler broadened line shape analysis : Variation of 'S' parameter showing a transition point, in the variation of electronic structure in the presence of variation of FA%.

7. Ratio of S-W parameter in ZnO-FA structure representing similar type of defects, introduced in the system, the grouping points near transition are encircled, but the exact transition point stands alone at the top.

8. Coincidence Doppler broadening (CDB) circuit diagram with two detectors, i)40 c.c HPGE solid state , ii) NaI (Tl) 3'' x3" scintillator detector kept at 180$^{o}$ opposite direction to receive the Two gamma rays at opposite direction.

9. Analysed results of CDB, ratio of counts in the region (511keV+ΔE) vs angular momentum distribution, with different FA% capped ZnO system : a) for valance band contributed region, that is detected with ZnO-FA, b) in the higher momentum region signifying the core electrons due to functional groups of FA , in ZnO-FA grains, using ZnO-pure as the standard spectrum.

10. FT-IR characteristic absorption frequencies for the as prepared Zn-FA nano grains with varying FA% used in the system .

**FIGURES :**

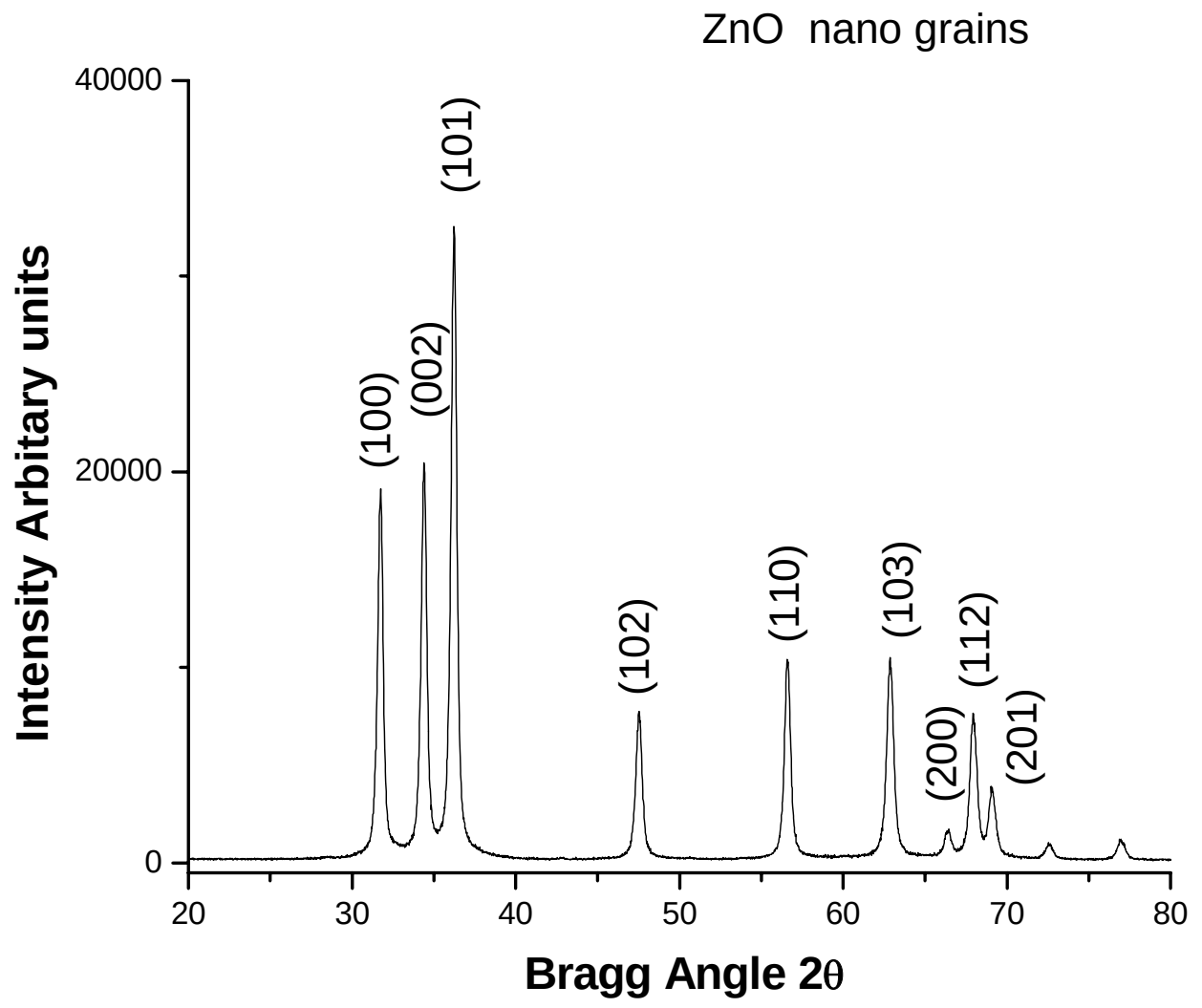

Figure 1.

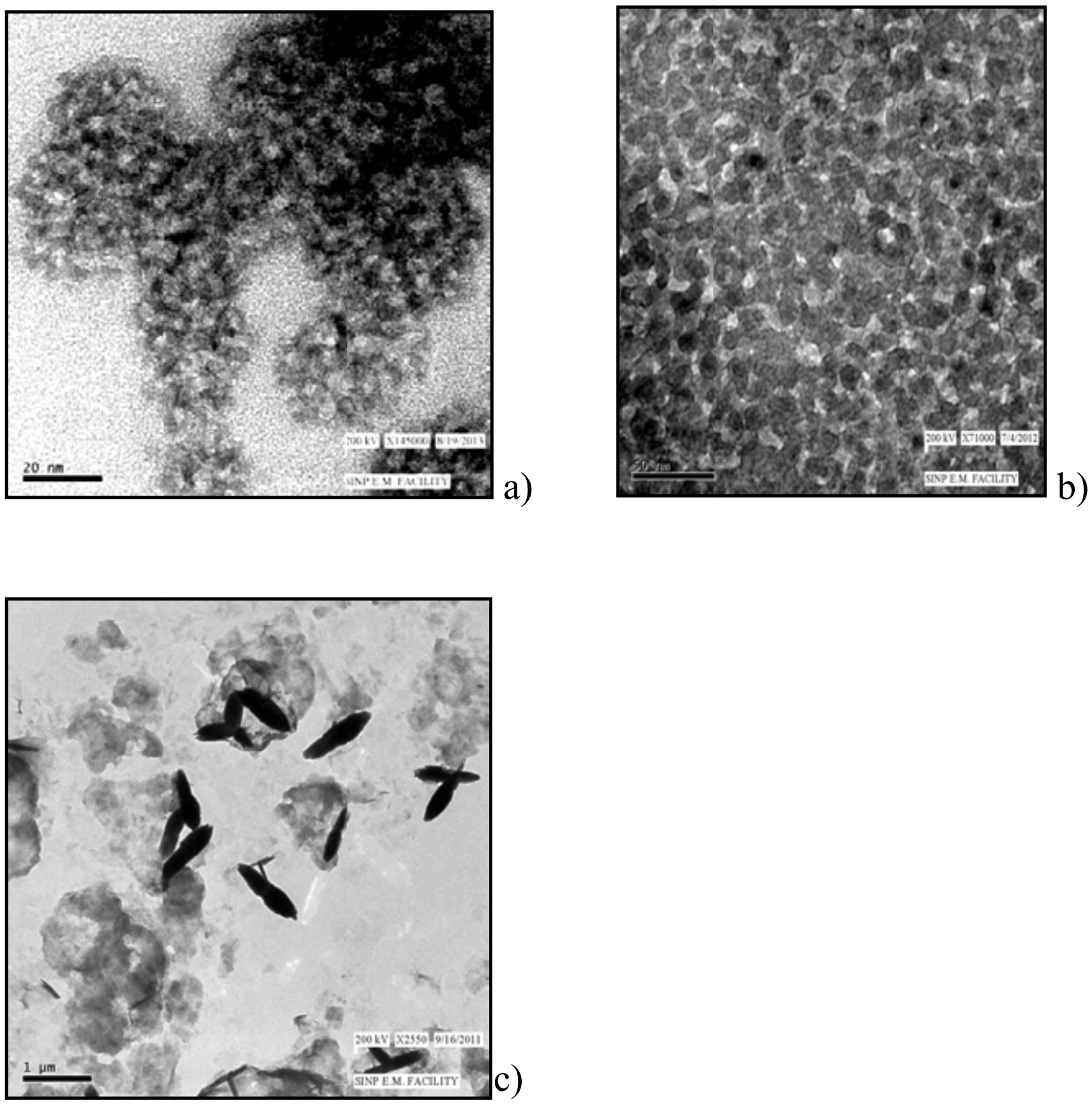
20 nm
200 kV X145000 8/19/2013
SINP E.M. FACILITY
a)
200 kV X71000 7/4/2012
SINP E.M. FACILITY
b)
1 µm
200 kV X2550 9/16/2011
SINP E.M. FACILITY
c)

Figure 2.

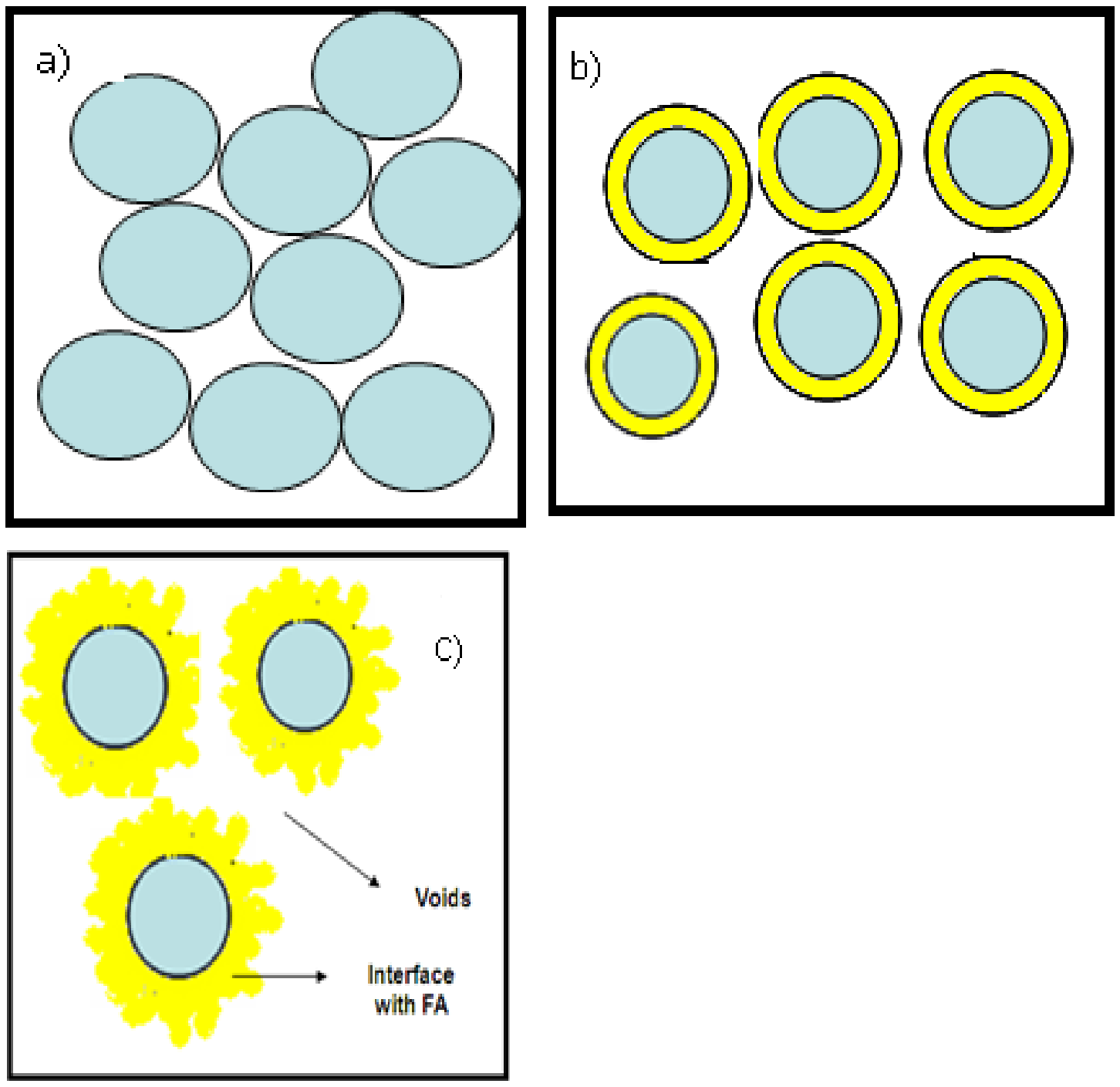

Figure 3.

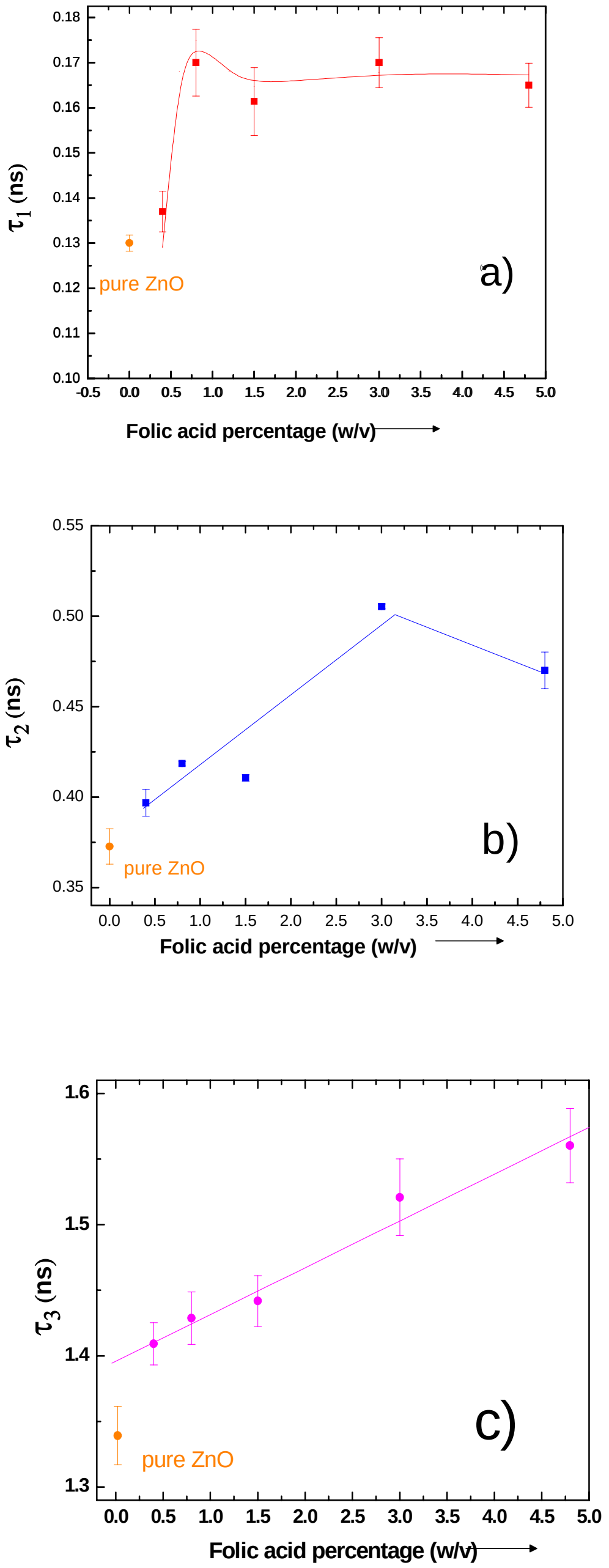
τ1 (ns)
pure ZnO
a)
Folic acid percentage (w/v)
τ2 (ns)
pure ZnO
b)
Folic acid percentage (w/v)
τ3 (ns)
pure ZnO
c)
Folic acid percentage (w/v)

Figure 4.

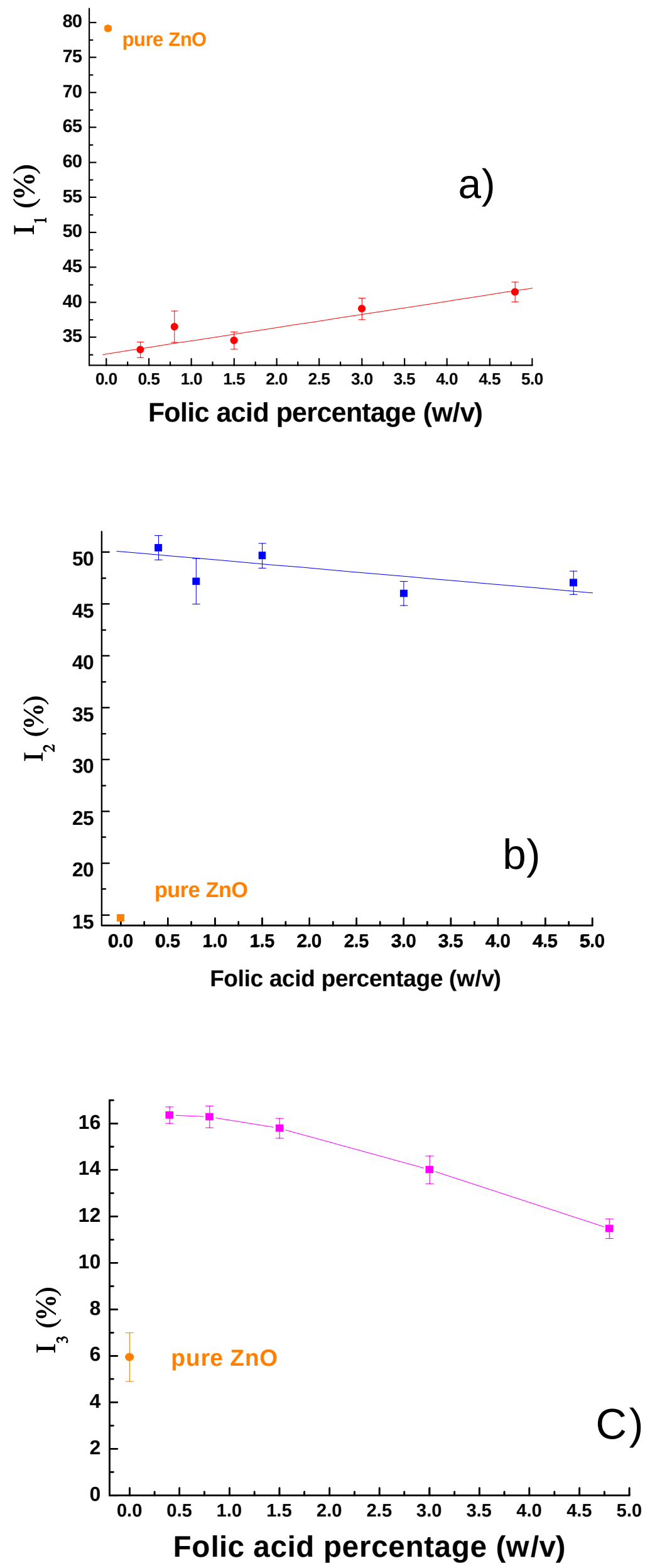
pure ZnO
a)
$I_1$ (%)
Folic acid percentage (w/v)
b)
pure ZnO
$I_2$ (%)
Folic acid percentage (w/v)
pure ZnO
C)
$I_3$ (%)
Folic acid percentage (w/v)

Figure 5.

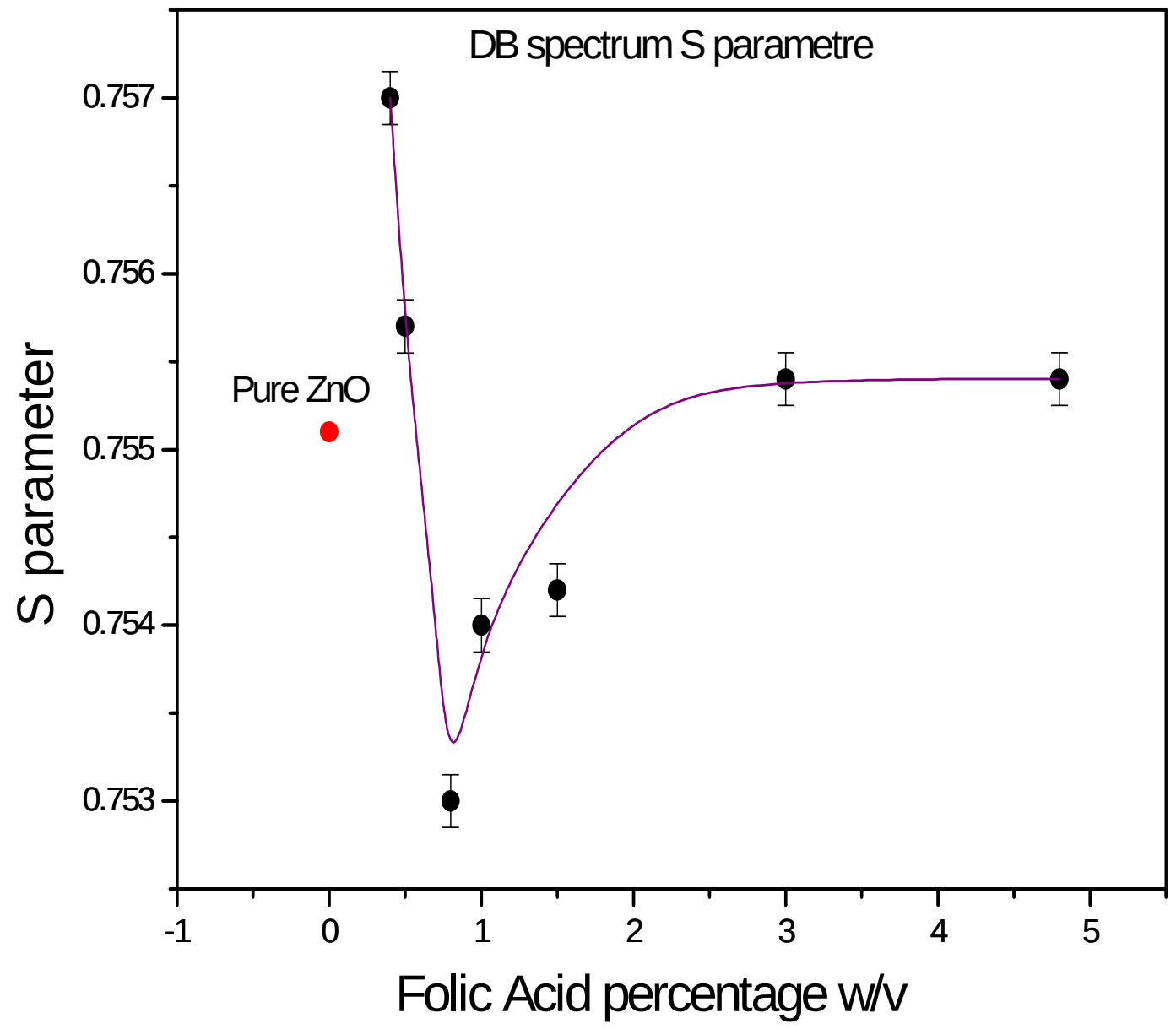

Figure 6.

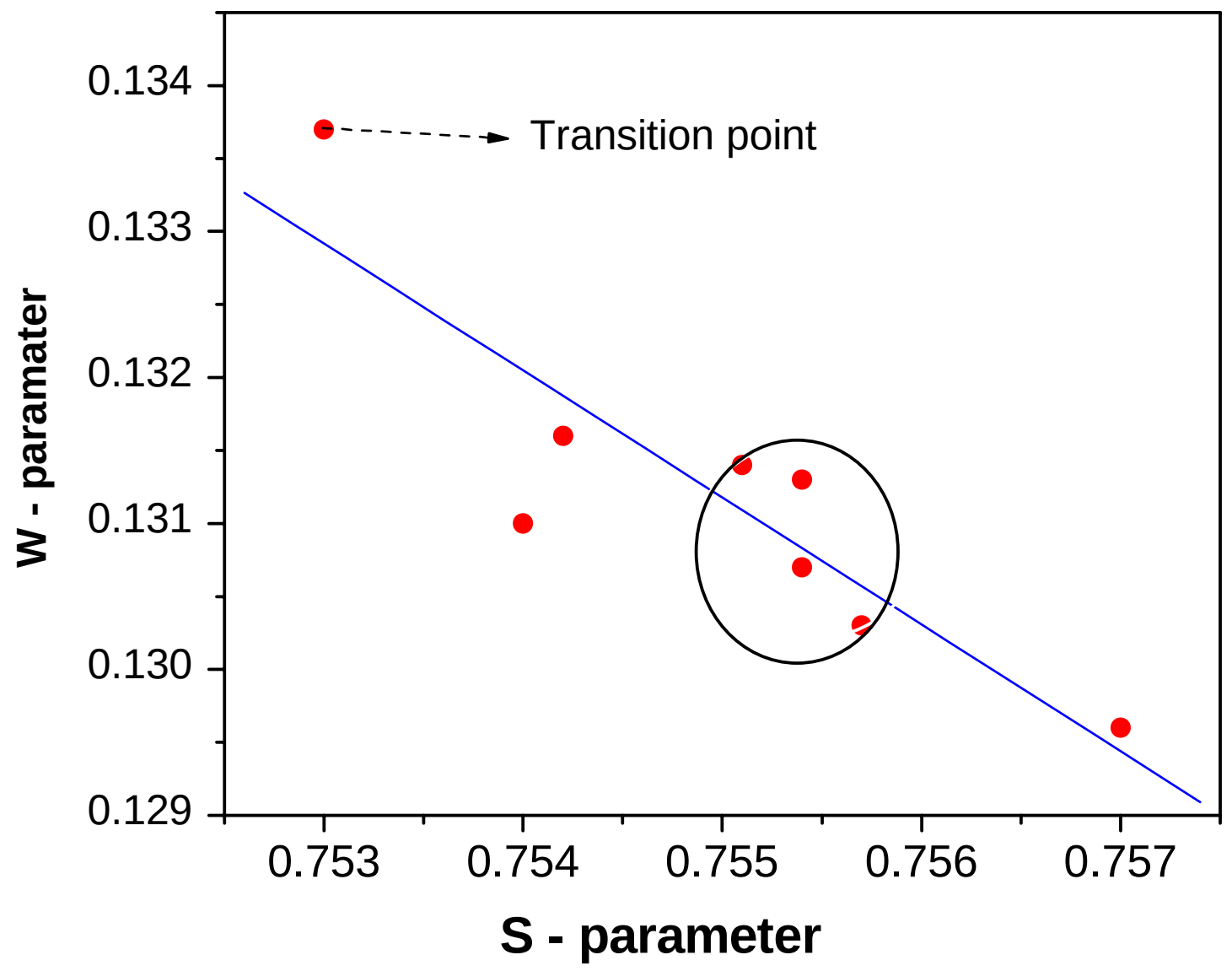

Figure 7.

Figure 8.

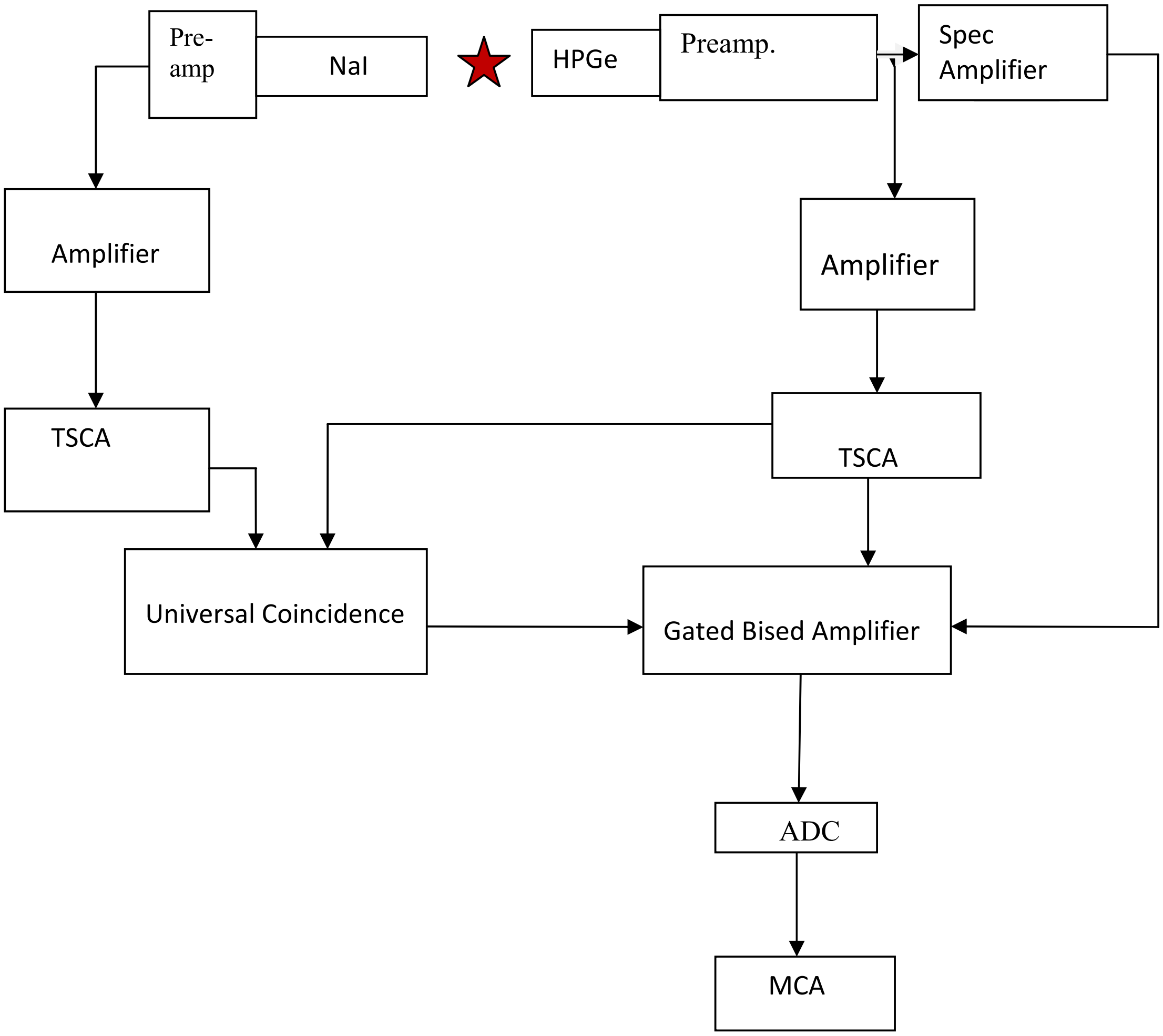

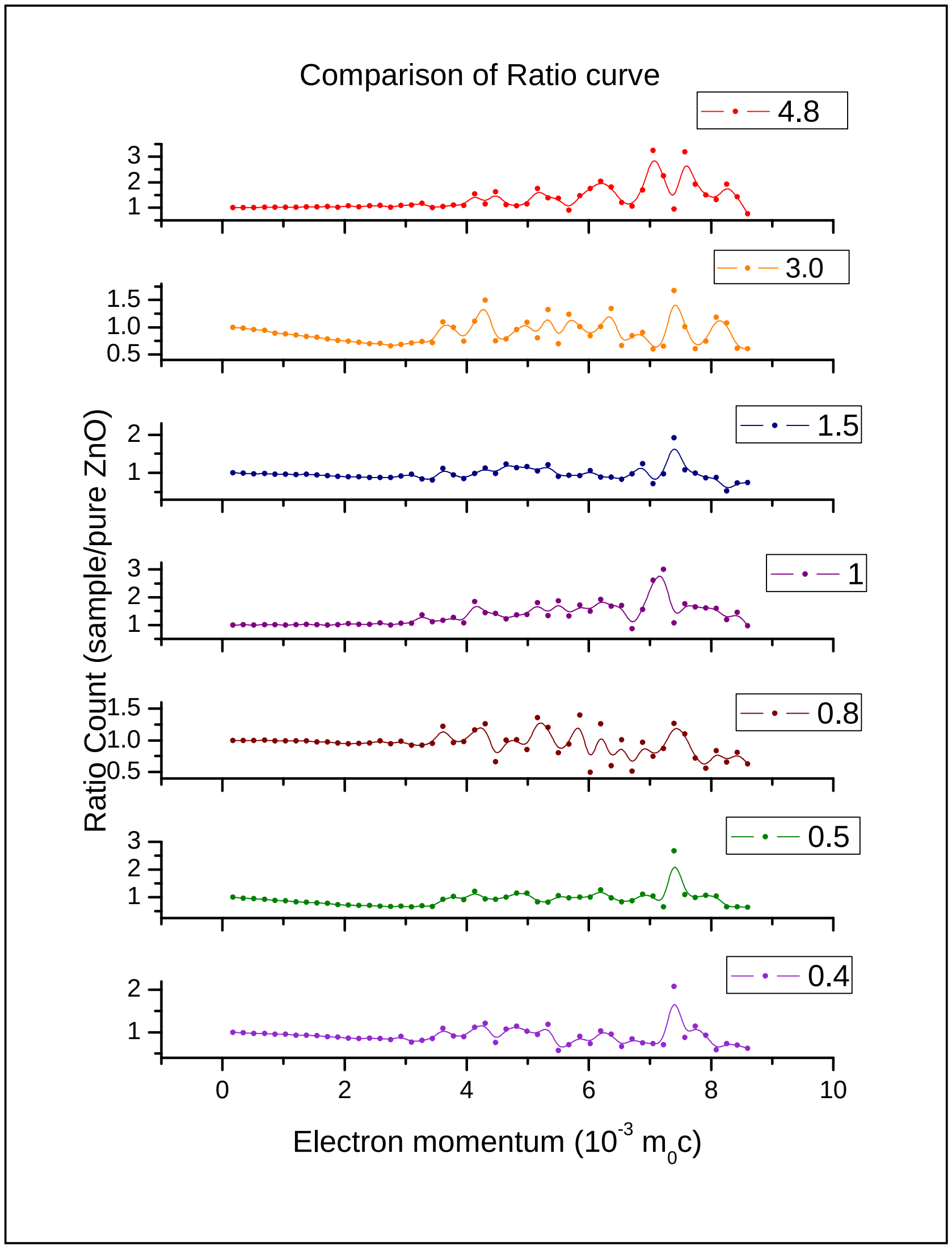

Comparison of Ratio curve
4.8
3.0
1.5
1
0.8
0.5
0.4
Ratio Count (sample/pure ZnO)
Electron momentum (10^-3 m0c)

Figure 9a

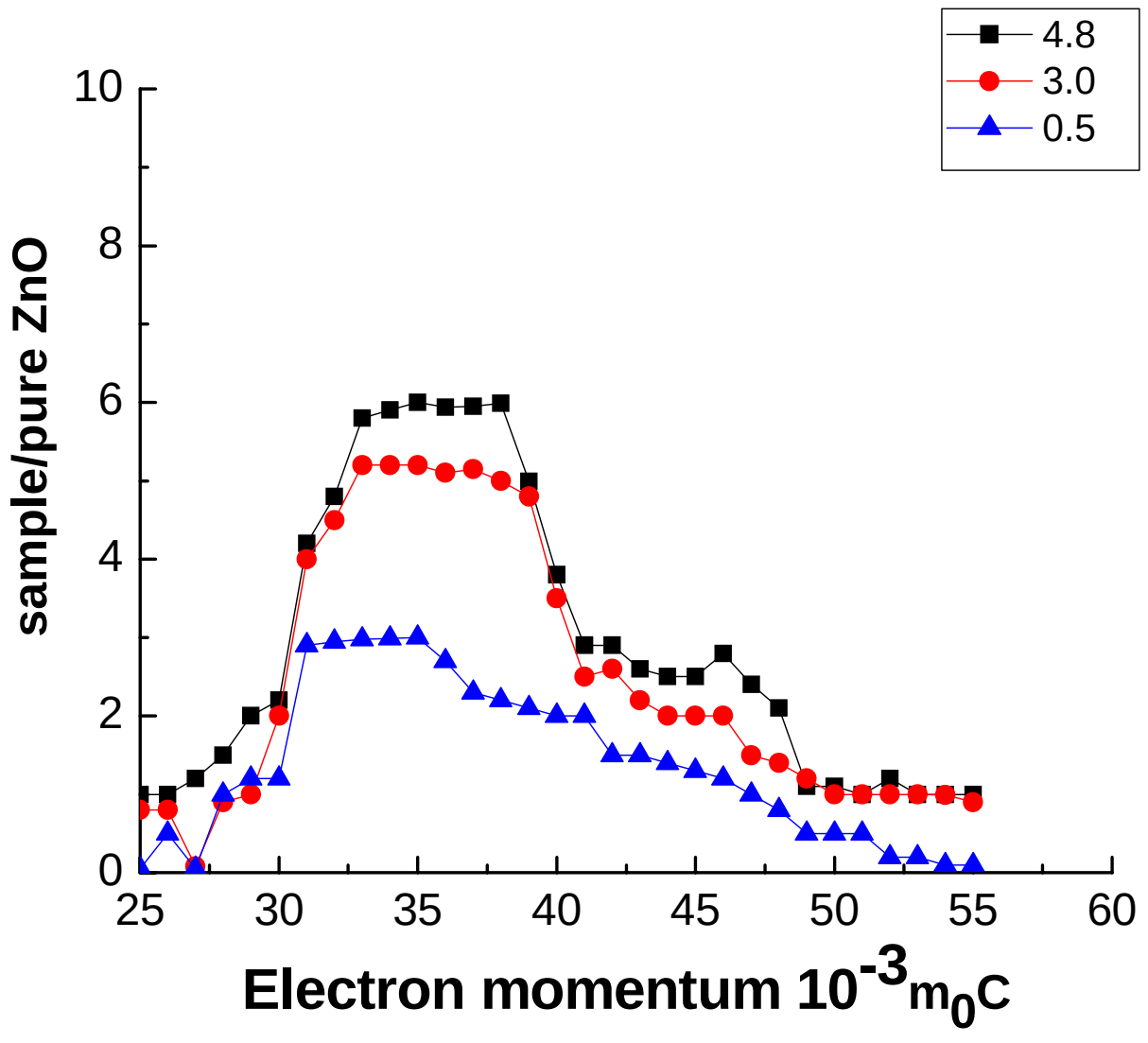

**Figure 9b**

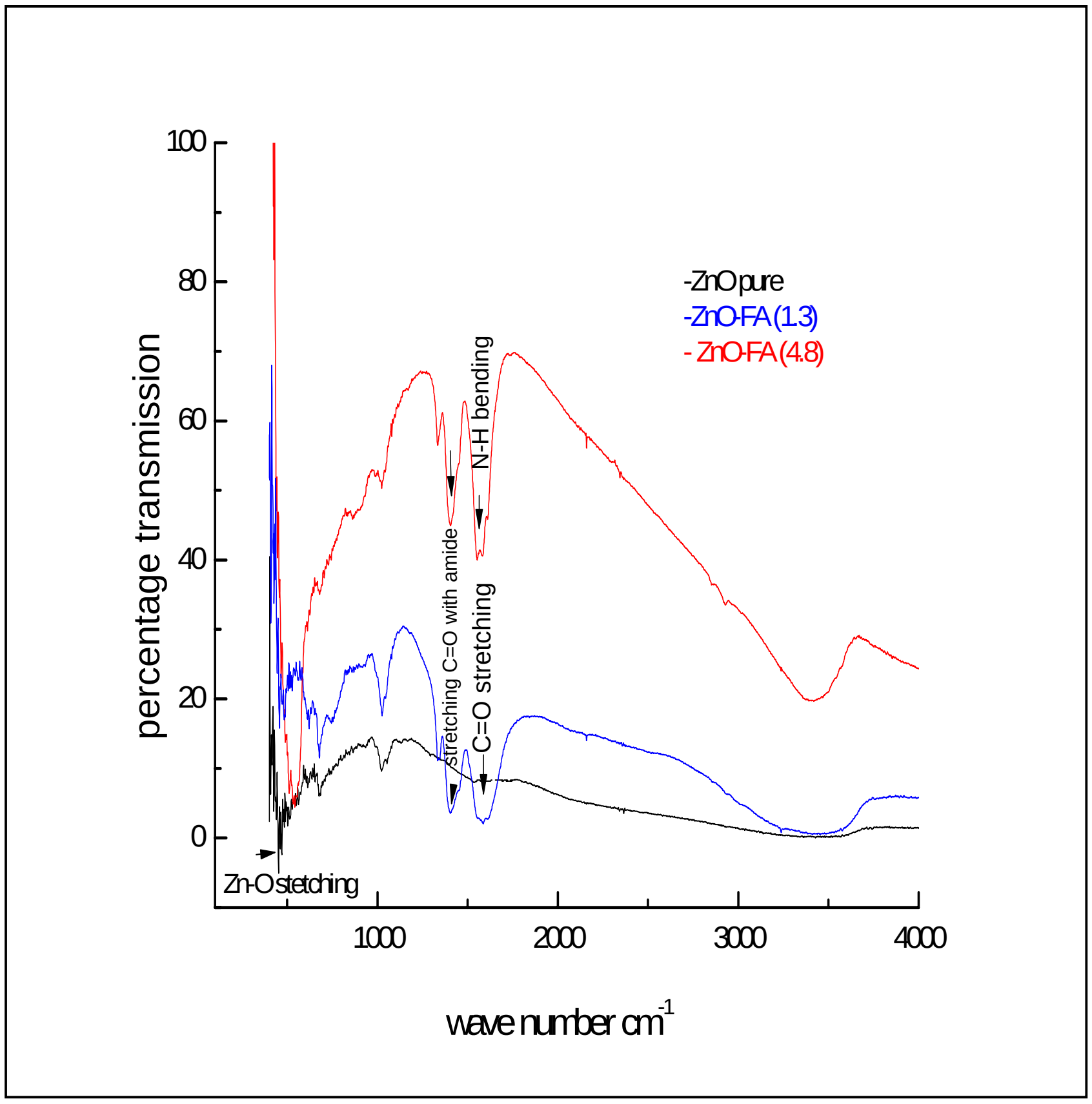

Figure 10.